\documentclass[12pt]{iopart}
\usepackage{latexsym}

\usepackage{latexsym}

\def\tr{{\rm tr}}
\def\ket#1{\mid~\!\!\!{#1}~\!\!\rangle}
\def\bra#1{\langle~\!\!{#1}~\!\!\!\mid}

\def\IF{if and only if }

\def\qm{quantum mechanics}

\def\cH{{\cal H}}

\def\cS{{\cal S}}

\def\${$\enskip }
\def\M{measurement }
\def\m{measurement}
\def\P{quantum probability law }

\begin{document}

\title[\bf Probability from minimal
measurement] {\bf Derivation of the
quantum probability law\\ from minimal
non-demolition measurement}

\author{F Herbut\footnote[1]{E-mail:
fedorh@infosky.net and
fedorh@mi.sanu.ac.yu}}

\address{Serbian Academy of
Sciences and Arts, Knez Mihajlova 35,
11000 Belgrade, Serbia}

\date{\today}

\begin{abstract}
One more derivation of the quantum
probability rule is presented in order
to shed more light on the versatile
aspects of this fundamental law. It is
shown that the change of state in
minimal quantum non-demolition
measurement, also known as ideal
measurement, implies the probability
law in a simple way. Namely, the very
requirement of minimal change of state,
put in proper mathematical form, gives
the well known L\"{u}ders formula,
which contains the probability rule.
\end{abstract}


\maketitle

\normalsize

\rm

\section{Introduction}

The quantum probability law \$\tr(E\rho
)\$ (its so-called trace-rule form) is
one of the fundamental pillars of
modern physics along with Einstein's
famous energy formula \$E=mc^2\$ and
Boltzmann's immortal entropy expression
\$S=klogW.\$ Gleason gave a seminal
derivation of the \P in his theorem
\cite{Gleason}. Nevertheless, as to
transparentness, there is much to be
desired. Though the \P looks simple,
there are "wheels within wheels" in it.
Therefore, it is important to view it
from as many different angles as
possible to be able to comprehend the
intricacies involved in it.

A number of alternative derivations
appeared in the literature. Let me
mention just a few.

(i) The approaches based on the
so-called eigenvalue-eigenstate link
\cite{Hartle}, \cite{Goldstone},
\cite{Squires};

(ii) The decision-theoretic approaches
\cite{Deutsch}, \cite{Barnum1},
\cite{Wallace}, \cite{Forrester};

(iii) Derivation from operational
assumptions \cite{Saunders};

(iv) The approach via entanglement.

The last mentioned approach went under
the title "Born's rule from envariance"
(environment assisted invariance).
There were 4 articles by Zurek
\cite{Zurek1}, \cite{Zurek2},
\cite{Zurek3}, \cite{Zurek4}, who
invented the approach, and there were 4
more articles by commentators
\cite{Schlossh}, \cite{Barnum2},
\cite{Mohrhoff}, \cite{Caves}, and
finally my own contribution in terms of
a complete theory of twin unitaries
(the other face of envariance)
\cite{FHtwinunit}. The first 8 articles
had two restrictions in establishing
essentially the trace rule \$\tr(E\rho
)\$ for probability, where \$E\$ was an
event (projector), and \$\rho\$ was the
subsystem density operator: they
handled only improper mixtures
\cite{d'Espagnat}, and did not go
beyond the commutation \$[E,\rho]=0$
restriction.

My article emphasized the role of
\$\sigma$-additivity in the derivations
from entanglement (the sole assumption
in Gleason's theorem). I suggested to
surmount the commutation restriction by
taking resort to minimal quantum
non-demolition (QND) measurement.

Subsequently I have realized that
minimal measurement is by itself
sufficient to derive the entire trace
rule. It has the advantage that it does
not require the \$\sigma$-additivity
assumption, and thus it is
complementary to Gleason's theorem
\cite{Gleason}. This article is devoted
to the exposition of the
minimal-measurement approach.

The paper is based on the idea that
{\it probabilities are predictions for
the statistical weights of
definite-result sub-ensembles in
measurement}. These are, in the end,
detected as relative frequencies.

\section{Assumptions of the Derivation}

We are dealing with an arbitrary
observable \$A\$ that has a {\it purely
discrete spectrum} \$\{a_n:\forall
n\}.\$ We write it in spectral form
$$A=\sum_na_nP_n,\quad n\not=
n'\enskip\Rightarrow a_n\not=
a_{n'}.\eqno{(1)}$$ It will be fixed
throughout. We have in mind QND {\it
\M} of the observable \$A$.\\

\subsection{The assumptions}

The {\it assumptions} of the approach
read as follows.

(i) {\it States are described by
density operators} \$\rho.\$

By "state" we mean an ensemble of
quantum systems prepared by a certain
procedure. Any measurement converts the
initial state \$\rho\$ into a final
state \$\rho'\$ (in the so-called
non-selective version, when the entire
ensemble is considered). The latter is
decomposable into states \$\rho_n'\$
that correspond to the different
results \$a_n\$ of \$A:\$
$$\rho'=\sum_nw_n \rho'_n;\quad \forall
n:\enskip w_n\geq
0,\enskip\sum_nw_n=1.\eqno{(2)}$$

If the \M is not a QND one, then the
states \$\{\rho_n':\forall n,w_n>0\}\$
need not be in any  simple relation to
\$A.\$ They correspond to definite
pointer positions on the measuring
instrument (which we make no use of in
this approach). The statistical weights
\$w_n\$ apply both to the states
\$\rho_n'\$ of the selective version
(in which definite results are
considered), and to the corresponding
pointer positions. {\it By the very
definition of \m, the weights equal the
probabilities}:
$$\forall n:\quad w_n=p(a_n,A,\rho
)\eqno{(3)}$$ (in obvious notation). In
other words, as it was stated in the
Introduction, the probabilities
\$p(a_n,A,\rho )\$ are understood to be
the predictions for the statistical
weights \$w_n,\$ which become relative
frequences when the measurement is
performed on the individual systems
that make up the ensemble.\\

{\it QND \m}, by definition, converts
an initial state \$\rho\$ into a final
state \$\rho',\$ which has two
properties:

(a) The states \$\rho_n'\$ that
determine the terms in decomposition
(2) {\it are dispersion-free with
respect to the observable} \$A$:
$$\forall n, w_n>0:\quad p(a_n,A,\rho_n')
=1.\eqno{(4)}$$

(b) If the initial state \$\rho\$ is
itself dispersion-free with respect to
\$A:\$ \$\exists n:\enskip p(a_n,A,\rho
)=1,\$ then so is the final state, and
the sharp value of \$A\$ is the same:
\$\rho'=\rho'_n,\$ but, in general, the
initial and the final states need not
be equal. (Earlier used synonyms for
"non-demolition" were "repeatable",
"predictive", "first-kind", etc.)\\

(ii) {\it Further, we assume that \IF a
state \$\rho\$ satisfies $$\tr(P_n\rho
)=1,\eqno{(5a)}$$ then the probability
\$p(a_n,A,\rho )\$ of the value \$a_n\$
of the observable \$A\$ in this state
is \$1$}. In other words, we assume the
validity of the trace rule for
probability-one events.

It is proved in Appendix A that (5a) is
(mathematically) {\it equivalent to}
$$P_n\rho P_n=\rho .\eqno{(5b)}$$

Let us denote by \$\rho_n''\$ {\it any
state} that has the sharp value \$a_n\$
of \$A:\$ \$p(a_n,A,\rho_n'')=1, \$and
let us consider the {\it family of all
mixtures}
$$\rho''\equiv
\sum_nv_n\rho_n'';\quad\forall
n:\enskip v_n\geq 0;\enskip\sum_nv_n=1.
\eqno{(6)}$$

An immediate consequence of (5b) is
that decomposition (6) can be rewritten
as
$$\rho''=\sum_nv_nP_n\rho_n''P_n,$$
which, on account of the orthogonality
and idempotency of the eigen-projectors
\$P_nP_{n'}=\delta_{n,n'}P_n,\$ implies
$$\rho''=
\sum_nP_n\rho''P_n.\eqno{(7)}$$ Since
(7) is obviously sufficient for (6),
also (7) characterizes states that are
mixtures of states with
definite values of \$A$.\\

If an initial state \$\rho\$ and an
observable (1) are given, then a {\it
subset} of the family of states (7) are
final states of QND measurements.\\

Our next-to-last {\it assumption} is:

(iii) {\it The state \$\bar\rho''\$ in
the family of states (7) that is {\bf
closest} to the initial state \$\rho\$
is the final state of a QND \M of the
observable \$A$.} By this, "closest" is
meant in the sense of minimal distance,
where distance is taken in the Hilbert
space \$\cH_{HS}\$ of all
Hilbert-Schmidt (HS) operators ( cf
\cite{RS} and Appendix B below). All
density operators are HS operators.

In general, also in a proper subset of
\$\cH_{HS},\$ in the set of all
trace-class operators, for which by
definition \$\tr\rho<\infty,\$ distance
is mathematically defined. We take
distance in \$\cH_{HS}\$ due to
Lemma-C in Appendix C.\\

Our last assumption is

(iv) {\it The probabilities
\$p(a_n,A,\rho )\$ are the same in all
\m s of \$A\$ in \$\rho$.}\\

\subsection{Discussion of the
assumptions}

Assumptions (i) and (iv) have a basic
(almost axiomatic) position in the
conceptual structure of \qm.

Assumption (ii) stipulates the trace
law for events that are certain. Here
we are on similar grounds as Zurek was
\cite{Zurek1}-\cite{Zurek4}, when he
set out to derive Born's rule assuming
its validity for events that are
certain. (In \cite{FHtwinunit} though,
when the full power of envariance was
made use of, the trace law under the
restriction \$[E,\rho ]=0\$ was derived
with no probability-law assumption to
start with.)

Assumption (iii) can be viewed as the
definition of minimal (or
minimal-disturbance) QND \m. Namely,
"closest" can be understood as
"minimally changed".

In the next section we derive
\$\bar\rho'',\$ and thus we obtain the
probabilities.\\

\section{Derivation of the trace rule}

We adapt now a former derivation
\cite{FH69} of the L\"{u}ders formula
\cite{Lud} to the present purpose.

The argument is very simple. It is
based on three almost evident
remarks:\\

{\bf Remark 1.} The super-operator
\$\hat P_A\equiv\sum_nP_n\dots P_n\$
(cf (1)) is a {\it projector} in
\$\cH_{HS}.\$  (The dots show the place
where any HS operator \$B\in\cH_{HS}\$
should be in the sum of products when
\$\hat P_A\$ is applied to it). One
easily shows the claimed Hermiticity
and idempotency of \$\hat P_A\$ in
\$\cH_{HS}\$ (cf Appendix B).

Let us denote by \$\cS_A\$ the subspace
of \$\cH_{HS}\$ onto
which \$\hat P_A\$ projects.\\

{\bf Remark 2.} As it is obvious from
(7), each density operator \$\rho''\$
from the family (6) (or (7)) is an
element of \$\cS_A$. And conversely,
the family (6) consists of all density
operators that are in \$\cS_A$.\\

{\bf Remark 3.} If \$\rho\$ is a
density operator, then so is its
projection \$\hat P_A(\rho )\$ (as easily
seen).\\

If \$\rho\$ is an arbitrary initial
state, its closest element in \$\cS_A\$
is its projection into \$\cS_A\$ (cf
Appendix D). The projection is a
density operator on account of Remark
3. The projection belongs to the family
(6) owing to Remark 2. Relation (3)
implies that the weights in the
projection give the probabilities.

Finally, let us write down the
projection. $$\hat P_A(\rho)=
\sum_nP_n\rho P_n.$$ This is the
well-known formula of L\"{u}ders, which
gives the change of state in minimal
QND measurement (also called ideal
measurement) \cite{Lud}.

Making the weights in the preceding
relation explicit, one obtains $$\hat
P_A(\rho)= \sum_n\Big(\tr(P_n\rho
)\Big) \Big(P_n\rho
P_n\Big/[\tr(P_n\rho )]
\Big).\eqno{(8)}$$

Relations (3) and (8) give our final
result: $$\forall \rho ,\enskip\forall
n:\quad p(a_n,A,\rho )=\tr(P_n\rho ).
\eqno{(9)}$$ In this way the trace-rule
form of the \P is derived.\\

Incidentally, if the event is
elementary (mathematically, a ray
projector)
\$P_n\equiv\ket{\phi}\bra{\phi}$, then
the \P is known in the form
\$\bra{\phi}\rho\ket{\phi}.\$ If also
the state is pure (mathematically also
a ray projector) \$\rho\equiv\ket{\psi}
\bra{\psi},\$ then one has the
transition-probability form
\$|\bra{\phi}\ket{\psi}|^2$. (All this
obviously follows from the
trace rule.)\\

{\bf Appendix A}

We prove now the following auxiliary
result that sheds light on assumption
(ii).

{\bf Lemma-A} If \$\rho\$ and \$P\$ are
a density operator and a projector
respectively, then \$\tr(\rho P)=1\$ is
{\it equivalent} to \$P\rho P=\rho\$ .

{\bf Proof.} It is obvious (by taking
the trace) that the latter relation
implies the former. Claim of the
inverse implication is not quite
trivial.

Since every density operator is a
trace-class operator, it has a finite
or countably infinite discrete positive
spectrum \$\{r_i:\forall i\}\$ (with
possible repetitions in the
eigenvalues). Hence, it can be written
in spectral form as
$$\rho =\sum_ir_i\ket{i}\bra{i},
\eqno{(A.1)}$$ where \$\ket{i}\$ is an
eigenvector corresponding to the
eigenvalue \$r_i.\$

The relation \$\tr(\rho P)=1\$ implies
\$\tr(\rho P^{\perp})=0\$
(\$P^{\perp}\equiv 1-P\$). Substituting
(A.1) in the latter relation, one
obtains \$\sum_ir_i\bra{i}P^{\perp}
\ket{i}=0.\$ On account of the
positivity \$\forall i:\enskip r_i>0,\$
and the easily seen non-negativity
\$\forall i:\enskip\bra{i}P^{\perp}
\ket{i}\geq 0,\$ one further has
\$\forall i:\enskip 0=\bra{i}P^{\perp}
\ket{i}= ||P^{\perp}\ket{i}||^2,\$ as
well as \$\forall i:\enskip
P^{\perp}\ket{i}=0,\$ and $\forall i:
\enskip P\ket{i}=\ket{i}.\$ Then,
applying \$P\dots P\$ to (A.1), one
obtains the second relation in Lemma-A.
\hfill $\Box$\\

{\bf Appendix B}

By definition, linear operators \$A\$
in a complex separable Hilbert space
are Hilbert-Schmidt ones if
\$\tr(A^{\dag}A)<\infty\$ (\$A^{\dag}\$
being the adjoint of \$A\$). The scalar
product in the Hilbert space
\$\cH_{HS}\$ of all linear
Hilbert-Schmidt operators is
\$\Big(A,B\Big)\equiv\tr(A^{\dag}B)\$
(cf the Definition after Theorem VI.21
and problem VI.48(a) in \cite{RS}).\\

{\bf Appendix C}

Let \$\cH\$ be a separable, complex
Hilbert space, and \$\cH_{HS}\$ the
Hilbert space of all linear
Hilbert-Schmidt operators in it (cf
Appendix B). Let, further,
\$\ket{\psi},\$ and \$\ket{\phi}\$ be
two arbitrary unit vectors in \$\cH.\$
The square of the {\it distance}
between them in \$\cH\$ is
$$\Big[d_{\cH}\Big(\ket{\psi},
\ket{\phi}\Big)\Big]^2\equiv
||\ket{\psi}-\ket{\phi}||^2=
\Big(\bra{\psi}-\bra{\phi}\Big)
\Big(\ket{\psi}-\ket{\phi}\Big)=
2-2Re\Big(\bra{\phi}\ket{\psi}\Big).
\eqno{(C.1)}$$ It depends on the
relative phase between the two vectors.

{\bf Definition-C} (i) We make the
convention that, whenever the distance
between two unit vectors in \$\cH\$ is
in question, it is understood that the
relative phase is chosen so that the
distance in (C.1) is minimal, i. e.,
that
$$\bra{\phi}\ket{\psi}\geq
0.\eqno{(C.2)}$$

(ii) We use the word "closer" in the
sense of "not farther", i. e., as
\$\leq ,\$ and not as \$<$.\\

{\bf Lemma-C} Let \$\ket{\psi},\$
\$\ket{\phi},\$ and \$\ket{\chi}\$ be
three arbitrary unit vectors in
\$\cH.\$ Then, taking the phase factors
of \$\ket{\phi}\$ and \$\ket{\chi}\$ in
accordance with Definition-C (i), the
former is closer than the latter to the
state vector \$\ket{\psi}\$ in \$\cH,\$
{\it \IF} the corresponding pure state
\$\ket{\phi}\bra{\phi}\$ is closer than
\$\ket{\chi}\bra{\chi}\$ to
\$\ket{\psi}\bra{\psi}\$ in
\$\cH_{HS}.\$ In other words, closer in
\$\cH\$ (observing Definition-C (i)) is
the case \IF it is true for the
corresponding ray projectors in
\$\cH_{HS}$.

{\bf Proof.} In view of (C.1) and
Definition-C (i), \$\ket{\phi}\$ is
closer to \$\ket{\psi}\$ than
\$\ket{\chi}\$ is to \$\ket{\psi}\$ \IF
$$\Big(2-2|\bra{\phi}
\ket{\psi}|\Big)\leq\Big(2-2|\bra{\chi}
\ket{\psi}|\Big)\enskip\Leftrightarrow
\enskip |\bra{\phi} \ket{\psi}|\geq
|\bra{\chi} \ket{\psi}|.\eqno{(C.3)}$$

On the other hand, one has
$$\Big[d_{\cH_{HS}}\Big(\ket{\psi}\bra{\psi},
\ket{\phi}\bra{\phi}\Big)\Big]^2=
\tr\Big[\Big(\ket{\psi}\bra{\psi}
-\ket{\phi}\bra{\phi}\Big)^2\Big]=
2-2|\bra{\phi}\ket{\psi}|^2.
\eqno{(C.4)}$$ Hence, the pure state
\$\ket{\phi}\bra{\phi}\$ is "closer" to
\$\ket{\psi}\bra{\psi}\$ than
\$\ket{\chi}\bra{\chi}\$ is to
\$\ket{\psi}\bra{\psi}\$ in
\$\cH_{HS}\$ \IF
$$|\bra{\phi}\ket{\psi}|^2\geq
|\bra{\chi}\ket{\psi}|^2.$$

Finally, since an inequality between
two non-negative numbers holds true \IF
the same inequality is valid between
their squares, one can see from (C.3)
and (C.4) that Lemma-C is proved.\hfill
$\Box$\\

{\bf Appendix D}

Now we prove (for completeness) a very
elementary auxiliary lemma.

{\bf Lemma-D} Let \$\cH\$ and \$\cS\$
be a separable (finite or infinite
dimensional) complex Hilbert space and
a subspace in it respectively. Let,
further, \$P\$ be the projector onto
\$\cS.\$ For every element \$a\in
\cH,\$ there is a unique element \$\bar
b\in\cS\$ that is closest to \$a\$
among all elements \$b\in\cH.\$ It is
\$\bar b\equiv Pa.\$ By this, "closest"
is meant in the sense of minimal
distance \$||a-b||$.

{\bf Proof.} For every \$a\in\cH ,\$
and every \$b\in\cS ,$ one can utilize
the orthogonality between the vectors
from the orthocomplement of \$\cS\$ and
those from \$\cS\$ itself:
$$||a-b||^2=
||(a-Pa)+(Pa-b)||^2=||a-Pa||^2+
||Pa-b||^2.$$ This is minimal with
respect to the choice of \$b\in\cS\$
\IF \$b\equiv Pa\$ because whenever
\$b\in\cS,\enskip b\not=Pa,\$
\$||Pa-b||^2>0$.\hfill $\Box$\\

\end{document}